\documentclass[applemac]{aa}
\usepackage{natbib}
\usepackage{txfonts}
\usepackage{graphicx}
\usepackage{color}
\bibpunct{(}{)}{;}{a}{}{,}

\begin{document}

\title{Impact of ice growth on the physical and chemical properties of dense cloud cores}
\subtitle{I. Monodisperse grains}
\author{O. Sipil\"a\inst{1}, P. Caselli\inst{1} \and M. Juvela\inst{2}
}
\institute{Max-Planck-Institut f\"ur Extraterrestrische Physik (MPE), Giessenbachstr. 1, 85748 Garching, Germany \\
e-mail: \texttt{osipila@mpe.mpg.de}
\and{Department of Physics, P.O. Box 64, 00014 University of Helsinki, Finland}
}

\date{Received / Accepted}

\abstract{We investigated the effect of time-dependent ice growth on dust grains on the opacity and hence on the dust temperature in a collapsing molecular cloud core, with the aim of quantifying the effect of the dust temperature variations on ice abundances as well as the evolution of the collapse. To perform the simulations, we employed a one-dimensional collapse model that self-consistently and time-dependently combines hydrodynamics with chemical and radiative transfer simulations. The dust opacity was updated on-the-fly based on the ice growth as a function of location in the core. The results of the fully dynamical model were compared against simulations run with different values of fixed ice thickness. We found that the ice thickness increases fast and reaches a saturation value (as a result of a balance between adsorption and desorption) of approximately 90 monolayers in the central core (volume density $\sim$$10^4\,\rm cm^{-3}$), and several tens of monolayers at a volume density of $\sim$$10^3\,\rm cm^{-3}$, after only a few $10^5\,\rm yr$ of evolution. The results thus exclude the adoption of thin ($\sim$10 monolayer) ices in molecular cloud simulations except at very short timescales. However, the differences in abundances and dust temperature between the fully dynamic simulation and those with fixed dust opacity are small; abundances change between the solutions generally within a factor of two only. The assumptions on the dust opacity do have an effect on the collapse dynamics through the influence of the photoelectric effect on the gas temperature, and the simulations take a different time to reach a common central density. This effect is however small as well. In conclusion, carrying out chemical simulations using a dust temperature corresponding to a fixed opacity seems to be a good approximation. Still, although at least in the present case its effect on the overall results is limited -- as long as the grains are monodisperse -- ice growth should be considered to obtain the most accurate representation of the collapse dynamics. We have found in a previous work that considering a grain size distribution leads to a complicated ice composition that depends non-linearly on the grain size. With this in mind, we will carry out a follow-up study where the influence of the grain size on the present simulation setup is investigated.}


\titlerunning{Impact of ice growth. I. Monodisperse grains}
\maketitle

\section{Introduction}

Dust grains play a key role in the star formation process (see \citealt{Oberg21} for an overview). While they have a direct effect on physical processes such as planet formation, they also have a great influence on chemical evolution across the many stages star formation. In molecular clouds, they provide a surface on which molecules can accrete and where many chemical reactions that are inefficient in the gas phase can readily proceed; a classic example of this being the formation of $\rm H_2$ \citep[e.g.,][]{Bron14}. This allows a rich chemistry to develop in the ices on the dust grains already in the cloud stage. The chemical complexity in the ices is then inherited at least partially by protostellar systems as they form insides cores located in the molecular clouds. Indeed, there is an increasing amount of evidence pointing to such an inheritance obtained from observations of Solar System objects, namely comets \citep[e.g.][]{Altwegg19,Drozdovskaya21}. As such, constraining ice chemistry in the cloud stage gives crucial clues to the initial conditions of forming planetary systems and finally the emergence of life.

In molecular clouds, where the temperature may be even lower than 10\,K deep inside cloud cores \citep[e.g.,][]{Crapsi07}, chemical evolution on grain surfaces is canonically thought to occur diffusively. In this picture, atoms and molecules arriving on the grain will scan the surface by moving from one binding site to the next by thermal hopping or quantum tunneling. Chemical reactions depend not only on the availability of the reactants but also on their mobility, meaning that reaction rates are diffusion-limited. A numerical description of this process, that has since been adopted in virtually all chemical models that treat grain-surface chemistry, was presented by \citet{Hasegawa92}, later expanded to include a treatment of layered ices \citep{Hasegawa93b}. Assuming that the diffusion proceeds thermally, the diffusion rate of a species $i$ is given by
\begin{equation}
R_{\rm diff}(i) = \nu_0 \exp(-E_{\rm b}(i)/T_{\rm dust}) \, ,
\end{equation}
where $\nu_0$ is the characteristic vibration frequency of the molecule on the surface, $E_{\rm b}(i)$ its binding energy, and $T_{\rm dust}$ is the dust temperature. The efficiency of chemical reactions is hence very strongly dependent on the dust temperature. This limitation is relaxed if the reactants are assumed to diffuse via quantum tunneling. However, the extent to which tunneling occurs is not yet fully known, and the dominating process is likely to depend on the type of surface. Experiments by \citet{Hama12} and \citet{Kuwahata15} have indicated that hydrogen and deuterium will mostly diffuse thermally on a water ice surface though a tunneling component is likely to be present, while within the ice the diffusion is expected to be dominated by thermal hopping along cracks in the ice \citep{Tsuge20}. Chemical models in the literature either (optionally) allow tunneling for hydrogen only \citep[e.g.,][]{Sipila15a,Vasyunin17} or neglect it \citep[e.g.,][]{Taquet12,Garrod13b,Ruaud16}. Even in the models where diffusion via tunneling is not allowed, barrier-mediated reactions may proceed via tunneling through the barrier, and this type of tunneling is not to be confused with the diffusion process. We note that nondiffusive processes have been recently invoked to explain the observed high abundances of complex organic molecules toward star-forming regions \citep{Shingledecker19,Jin20, Garrod22}. These too depend to some extent on $T_{\rm dust}$ because diffusive reactions are included as potential initiating reactions for the nondiffusive chemistry.

It is clear from the above that accurate estimations of $T_{\rm dust}$ are required so that reaction rates on grain surfaces can be estimated reliably. However, $T_{\rm dust}$ may play a role in the gravitational collapse of cloud cores as well. This is because the gas temperature is regulated by collisional coupling with dust grains at densities exceeding $\sim$$10^5 \, \rm cm^{-3}$ \citep{Goldsmith01} and hence the thermal balance will receive a contribution from $T_{\rm dust}$. The value of $T_{\rm dust}$ at each point in the core is time-dependent because of the continual accretion of molecules onto grains, which increases the ice thickness and thus modifies the dust opacity. Indeed, opacity variations within single objects have been observed (see for example \citealt{Chacon-Tanarro19b} and references therein), though we note that it is not clear in the general case whether the observed variations are due to ice thickness variations or grain growth, or possibly both. Nevertheless it is not straightforward to deduce if and how the opacity variations can influence core collapse.

There exist many numerical models of star-forming regions that consider the influence of ice opacity on $T_{\rm dust}$. Some examples of these are \citet{Zucconi01,Keto08} for starless and prestellar cores, \citet{Kamp18,Ballering21,Arabhavi22} for protostellar cores, and the work of \citet{Hocuk17} who presented parameterized relations applicable in several types of interstellar environment. The aforementioned works treat the connection between chemistry and dust opacity in various ways, in a mostly time-independent fashion. To the best of our knowledge, simulations of collapsing clouds that incorporate radiative transfer-based simulations of $T_{\rm dust}$ connected with time-dependent chemistry and dust opacity have never been performed. The presentation of such a model is the main aim of the present paper. By running fully time-dependent simulations, we can provide estimates of how dust opacity variations affect the chemical and physical evolution of a collapsing cloud core.

The paper is structured as follows. In Sect.\,\ref{s:model}, we discuss the setup of our numerical model and its various components. We present the results of our simulations in Sect.\,\ref{s:results} and discuss them in further detail in Sect.\,\ref{s:discussion}. We give our conclusions in Sect.\,\ref{s:conclusions}. Some additional results and discussion are provided in Appendix~\ref{appendixA}.

\section{Model}\label{s:model}

We employ the hydrodynamical collapse model HDCRT introduced in our two previous papers \citep{Sipila18,Sipila22}, where its functionality is described in detail. Briefly, the model is one-dimensional (1D) and uses the Lagrangian approach to follow the evolution of tracer particles. The model combines the hydrodynamics equations with chemical and radiative transfer simulations so that all relevant quantities such as the chemical abundances or line cooling powers are all calculated self-consistently and time-dependently as the simulation is progressing. We use very extensive chemical networks, as opposed to post-processing the chemistry after the hydrodynamical simulation has been carried out using a simple chemical network, to ensure that the evolution of the various molecules is captured as accurately as possible. Thanks to various optimizations \citep[see][]{Sipila22}, a workflow designed with parallel computations in mind, and the use of a 1D geometry, the simulations can be run in a reasonable timescale even with a large chemical network. Running a full collapse simulation with the present setup (see below for additional details) takes less than a week on a personal computer.

In the past works \citep{Sipila18,Sipila22} we employed a very extensive chemical network including deuterium and spin-state chemistry, with $\sim$1000 chemical species connected by over 40,000 reactions, so that we could investigate the evolution of deuterium-bearing tracer molecules at high volume densities where depletion onto grain surfaces is efficient. Here we simplify the chemistry by taking deuterium and spin states out of the chemical networks so that we may run the simulations faster\footnote{With deuterium and spin state included, the simulations take a factor of a few more time to run, that is, two to three weeks on a personal computer as opposed to five to six days for networks without them.}. Essentially, our chemistry setup corresponds to the 2014 version of the public KIDA gas-phase network (kida.uva.2014; \citealt{Wakelam15}) and a grain-surface network including reactions collected from several sources, mainly \citet{Semenov10}. The grain-surface network is however rather limited, and we can at present estimate reliably the grain-surface evolution of molecules up to methanol in complexity.

For this paper, we have extended HDCRT to take into account the variations in dust properties across the simulated cloud. New dust models are created on the fly based on the ice abundances at each time step (see also below). For this, we employ the {\tt optool} program \citep{Dominik21}, which allows the user to construct a custom dust model based on a set of input parameters and gives as standard output the absorption and scattering opacities as a function of wavelength. Here, we assume that the grains consist of a carbon/silicate core in a 40/60 mass ratio\footnote{The value of the ratio is associated with uncertainty; for example, ratios of $\sim$28/72 and $\sim$20/80 have been proposed by \citet{Draine11} and \citet{Tielens05}, respectively.} and that the grain radius is 0.1\,$\rm \mu m$, which yields an average grain material density of $\rho_{\rm d} = 2.54\,\rm g \, cm^{-3}$ for the grain core. We use the same value in our chemical simulations. The refractive indices for the graphite and silicate grain material are taken from \citet{Zubko96} and \citet{Draine03}, respectively.

We consider various levels of coating of the grains by water ice. In our full simulation, the mass fraction of the ice relative to the grain core varies as a function of location in the cloud with the accretion of gas-phase material onto the grains. This is treated by running {\tt optool} separately for each chemical tracer cell in the hydrodynamical simulation\footnote{The hydrodynamical simulation consists of 1000 tracer cells, initially spaced evenly throughout the core. For the chemistry and radiative transfer, we use a subset of 35 cells, spaced such that $\sim$60\% of the cells lie inside the inner fourth of the core in terms of radius, and the rest are distributed in the outer core with increasing sparsity. The use of low resolution in the outer core is justified because the chemical abundance gradients are shallow; the limited overall resolution also helps the required computational time to stay within acceptable limits. The resolution of the hydrodynamical simulation needs to be high, however, in order to obtain an accurate solution \citep[see Table~3 and related discussion in][]{Sipila18}.}, taking the local mass fraction of the ice (which depends on the sum of the number densities of all ice species) at a given simulation time as an additional input for {\tt optool}. Even though in reality the ice consists of a wide variety of species, we assume for the purposes of the opacity calculation that the ice is made up of water only, with refractive index data taken from \citet{Warren08}. We note that there is a small discrepancy between the ice opacity calculations and the chemical simulations, which assume a time-independent grain radius of 0.1\,$\rm \mu m$, that is, the growth of the ice is not taken into account. The discrepancy is small enough that we consider it passable to neglect its effect on the chemistry. A fully consistent model could be built by changing the dust-to-gas mass ratio locally as a function of chemical evolution -- without grain-grain collisions, the grain number density remains constant at all times.

As noted in the Introduction, there exist to our knowledge no preceding collapse models that have adopted time-independent dust opacities. To better assess the impact of the dynamic approach to dust evolution on the simulation results, we have in addition to the full simulation run three additional simulations where the ice mass fraction is set to a constant value corresponding to 80, 10, or 0 monolayers of water ice (i.e., bare grains in the last case). In this way, we can determine the effect of the (evolving) dust properties on the collapse duration and on the abundances of various molecules as a function of time, in particular those that are produced via barrier-mediated reactions on grain surfaces, such as methanol. The simulations along with their designations are compiled in Table~\ref{tab:simulationParameters}.

\begin{table}
\begin{center}
\caption{Simulations run for this paper.}
\begin{tabular}{ll}
\hline
\hline
Simulation label & Explanation \\
\hline
M\_TD &  time-dependent dust models constructed\\
  & on the basis of the local ice thickness\\
M\_80 &  fixed dust model corresponding to 80 \\
  & monolayers of ice on the grains\\
M\_10 & fixed dust model corresponding to 10\\
 &  monolayers of ice on the grains\\
M\_0 &  fixed dust model corresponding to\\
&  bare grains\\
\hline
\label{tab:simulationParameters}
\end{tabular}
\end{center}
\end{table}

We calculate $T_{\rm dust}$ using the Monte Carlo radiative transfer code CRT \citep{Juvela05}. Given the current density distribution and the adopted dust properties, CRT uses a large number of photon packages to simulate how photons from the external radiation field are scattered and absorbed inside the cloud core. The resulting knowledge of the absorbed energy is used to calculate the equilibrium dust temperatures for each cell. The program allows the inclusion of multiple dust populations, which are taken into account both during the radiative transfer simulation and when the dust equilibrium temperatures are solved. In practice, the dynamic dust opacities are implemented by giving each dust model a relative abundance of 1 in the cell with the corresponding ice thickness, and a relative abundance of 0 elsewhere. The calculation of $T_{\rm dust}$ via radiative transfer means is a computationally intensive process. However, according to our tests performed with CRT, the impact of taking several dozen dust components in place of one has an almost negligible effect on the computation time, though this result may vary with the radiative transfer code being used. Constructing the dust models with {\tt optool} only takes about one second of total computation time per cell owing to the simplicity of the model (spherical monodisperse grains), and this combined with the fact that the dust temperature is not updated at every hydrodynamical time step \citep[see][]{Sipila22} means that the overall computational cost of the dust opacity model creation is small compared to the overall duration of the simulation. Here we of course also benefit from the use of a 1D physical model which limits the number of cells where the radiative transfer simulations need to be performed. To introduce time savings in more complex physical model setups, one could precalculate the dust models with a fine grid of ice thicknesses, and then use the dust model closest to the actual ice thickness in each model cell at any given simulation time. This would introduce some error, however, and with our present simulation setup such time savings are not necessary.

The dust temperature and its temporal evolution has an impact on the molecular abundances in the ice, and possibly on the duration of the final stages of the collapse, as outlined in the Introduction. We have shown in \citet{Sipila22} that chemistry plays a very minor role for the dynamics of the late stages of the collapse because the effect of line cooling is limited once the collapse is well underway. Here we wish to constrain the magnitude of the effect due to the variations in $T_{\rm dust}$ caused by ice growth; effects may arise not only in the inner core where the dust-gas coupling is strong, but also in the outer core because the dust opacity affects the efficiency of photoelectric heating. Our goal is to study these points on the general level, as opposed to attempting to reproduce observations toward particular targets, and in keeping with this goal we have taken for the initial cloud a generic Bonnor-Ebert sphere \citep{Bonnor56, Ebert55} with a mass of $20\,M_{\odot}$ and an initial temperature $T_{\rm gas} = 20\,\rm K$. The initial central density of the cloud is $n({\rm H_2}) = 10^4 \, \rm cm^{-3}$ and the non-dimensional radius, which controls the stability of the cloud, is set to 10 (the outer radius in physical units is $\sim$99,000 au). These choices ensure that the initial cloud is in a supercritical configuration and is thus expected to start to (slowly) collapse already at the beginning of the hydrodynamical simulation. In contrast to \citet{Sipila22}, we assume an initial temperature of 20\,K as opposed to 10\,K because the former value better represents the equilibrium temperature in the outer core -- once the simulation starts, the gas temperature in the inner core starts decreasing rapidly towards 10\,K. To compensate for the higher temperature, the core mass and nondimensional radius were scaled compared to \citet{Sipila22} (where we assumed values of $10\,M_{\odot}$ and 23, respectively) so that an initially supercritical configuration could still be attained.

\begin{table}
        \centering
        \caption{Initial abundances (with respect to the total hydrogen number density) used in the chemical modeling.}
        \begin{tabular}{l|l}
                \hline
                \hline
                Species & Abundance\\
                \hline
                $\rm H_2$ & $5.00\times10^{-1}$\\
                $\rm He$ & $9.00\times10^{-2}$\\
                $\rm C^+$ & $1.20\times10^{-4}$\\
                $\rm N$ & $7.60\times10^{-5}$\\
                $\rm O$ & $2.56\times10^{-4}$\\
                $\rm S^+$ & $8.00\times10^{-8}$\\
                $\rm Si^+$ & $8.00\times10^{-9}$\\
                $\rm Na^+$ & $2.00\times10^{-9}$\\
                $\rm Mg^+$ & $7.00\times10^{-9}$\\
                $\rm Fe^+$ & $3.00\times10^{-9}$\\
                $\rm P^+$ & $2.00\times10^{-10}$\\
                $\rm Cl^+$ & $1.00\times10^{-9}$\\
                \hline
        \end{tabular}
        \label{tab:initialAbundances}
\end{table}

\begin{table}
        \centering
        \caption{Values of the physical parameters kept fixed in all simulations.}
        \begin{tabular}{lc}
                \hline
                \hline
                Parameter & Value\\
                \hline
                cosmic ray ionization rate ($\zeta$) & $1.3\times10^{-17} \, \rm s^{-1}$\\
                grain radius ($a_{\rm g}$) & $10^{-5} \, \rm cm$\\
                grain material density ($\rho_{\rm d}$) & $2.54 \, \rm g \, \rm cm^{-3}$\\
                diffusion-to-binding energy ratio ($E_{\rm d}/E_{\rm b})$ & $0.6$\\
                dust-to-gas mass ratio ($R_{\rm d}$) & $0.01$\\
                visual extinction at the core boundary ($A_{\rm V}$) & 1\,mag\\
                scaling of the ultraviolet field ($G_0$) & 1\\
                \hline
        \end{tabular}
        \label{tab:physicalParameters}
\end{table}

The initial abundances of the chemical simulations are essentially the same as in \citet{Sipila22} (with the exception that here we do not consider deuterium or spin-state chemistry) and are reproduced in Table~\ref{tab:initialAbundances}; the results of a test simulation where the initial hydrogen content is distributed between atomic and molecular hydrogen are described in Appendix~\ref{appendixB}. We perform chemical simulations using our gas-grain chemical model {\sl pyRate}, which couples gas-phase and grain-surface chemistry via adsorption and desorption. The general functionality of the model is described in \citet{Sipila15a}, while subsequent updates are presented in \citet{Sipila19a}. The model includes several (non-thermal) desorption mechanisms: chemical desorption under the assumption that 1\,\% of the products of exothermic grain-surface reactions leave the grain surface \citep{Garrod07}; photodesorption of CO, $\rm H_2O$, $\rm CO_2$, and $\rm N_2$ with yields taken from \citet{Oberg09a,Oberg09b}; and cosmic ray induced desorption using the revised method presented in \citet{Sipila21}. We assume that diffusion on the grain surface occurs thermally.

The values of various physical parameters fixed in the simulations are collected in Table~\ref{tab:physicalParameters}. The interstellar radiation field spectrum is taken from \citet{Black94}, and we employ the standard unscaled ultraviolet field (i.e., $G_0 = 1$) in calculations of photoreaction rates whose rate coefficient follows the formula $G_0\,e^{(-\gamma A_{\rm V})}$, where the value of $\gamma$ depends on the reaction. The model includes self-shielding of $\rm H_2$, CO, and $\rm N_2$. $\rm H_2$ is treated following \citet{Draine96}. The self-shielding factors of CO are taken from \citet{Visser09} and those of $\rm N_2$ from \citet{Li13} \& \citet{Heays14}.

\section{Results}\label{s:results}

\subsection{Time-dependent physical structure}

\begin{figure*}
\centering
        \includegraphics[width=2.0\columnwidth]{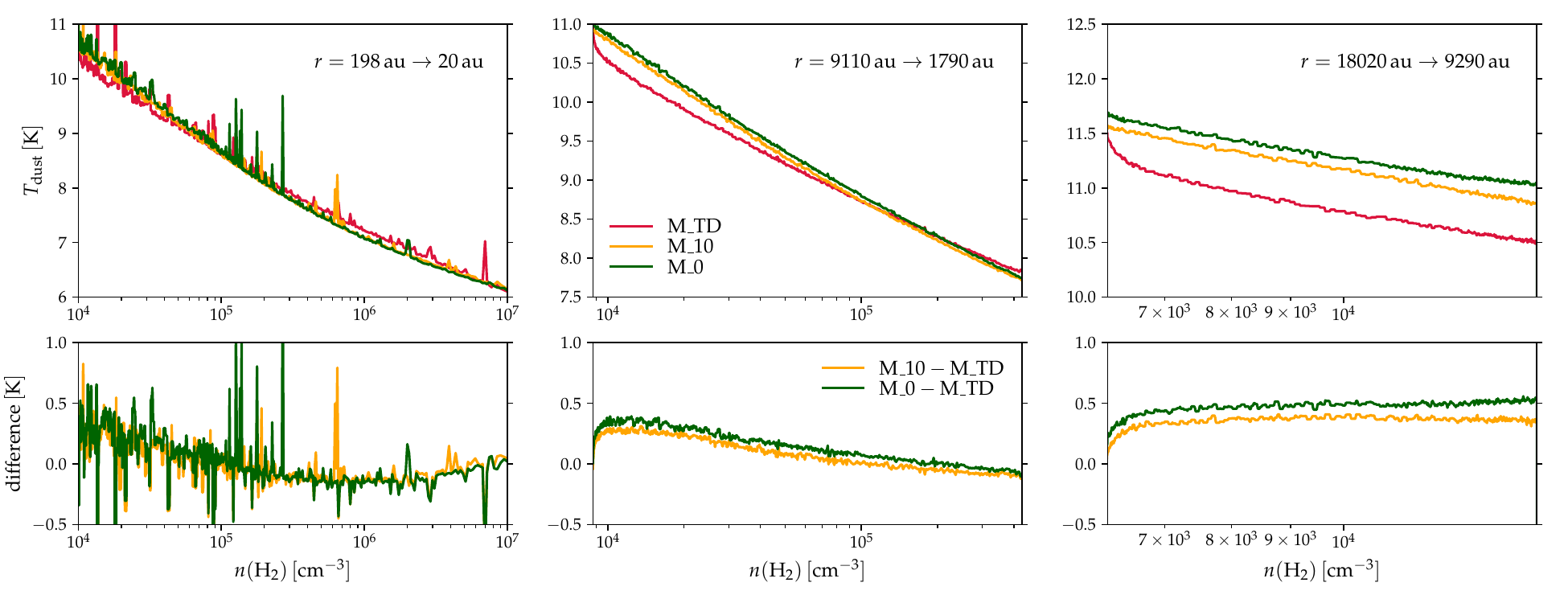}
    \caption{Evolution of $T_{\rm dust}$ as a function of volume density in three model cells in simulations~M\_TD, M\_10, and M\_0 (upper panels). The volume density in each cell increases as a function of time due to infall, and hence the time advances to the right. The time itself is not shown because the simulations evolve at different rates (it takes a different amount of time to reach a certain volume density) and hence a common reference time cannot be defined easily. Each cell starts out at a different location in the core; the initial and final location are given in the upper right corner in each panel. The quoted values correspond to simulation~M\_TD. The corresponding locations in the other two simulations deviate by less than ten per cent from these values despite the different overall evolution of the cores. The lower panels show the difference of the temperature curves as given in the middle bottom panel.}
        \label{fig:physEvoManyCells}
\end{figure*}

\begin{figure*}
\centering
        \includegraphics[width=2.0\columnwidth]{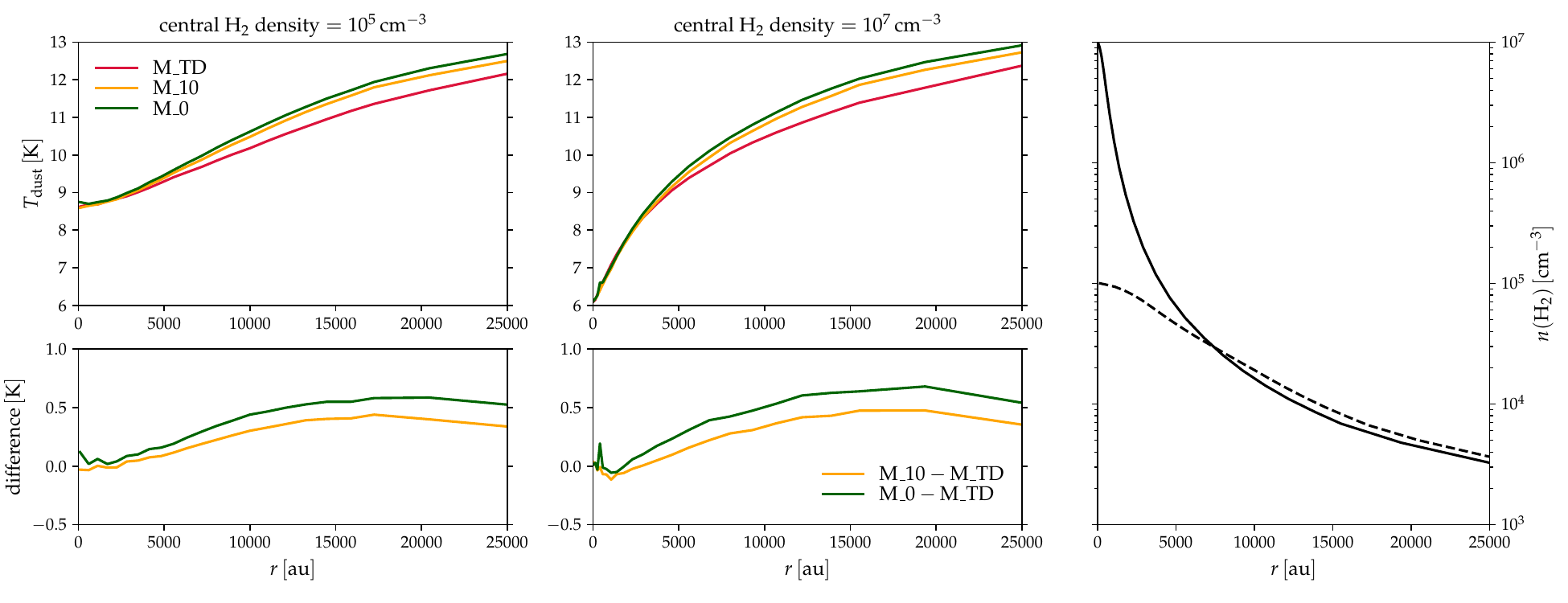}
    \caption{Snapshots of the dust temperature as a function of radius in simulations~M\_TD, M\_10, and M\_0 at the time when the central density of the core hits either $n({\rm H_2}) = 10^5 \, \rm cm^{-3}$ (upper left-hand panel) or $n({\rm H_2}) = 10^7 \, \rm cm^{-3}$ (upper middle panel). As in Fig.\,\ref{fig:physEvoManyCells}, the lower panels show the difference between the temperatures for clarity. The right-hand panel displays the volume density profile at the two central density values in simulation~M\_TD; the corresponding profiles in simulations~M\_10 and M\_0 are only marginally different to those shown here.}
        \label{fig:physEvoTime}
\end{figure*}

\begin{table}
\begin{center}
\caption{Evolutionary time to reach a central density of $n({\rm H_2}) = 10^7 \, \rm cm^{-3}$ in each simulation.}
\begin{tabular}{ll}
\hline
\hline
Simulation label & Time\,[yr] \\
\hline
M\_TD &  $7.00 \times 10^5$\\
M\_80 & $6.40 \times 10^5$ \\
M\_10 & $7.11 \times 10^5$\\
M\_0 &  $7.38 \times 10^5$\\
\hline
\label{tab:evoTimes}
\end{tabular}
\end{center}
\end{table}

A comparison of the results of the four simulations shows that the choice of dust properties does have an effect on the collapse duration, and the four simulations take a different amount of evolutionary time to reach the termination condition -- the simulations are designed to end when the infall velocity exceeds 1.5 times the local sound speed at any point in the core. The termination condition is however arbitrary, and it is in the present case more instructive to compare the progress of the simulations up to a common point in the evolution. We therefore investigate the evolution of the model cores up to the evolutionary time when a central density of $n({\rm H_2}) = 10^7 \, \rm cm^{-3}$ is reached. The time required to reach this density in each simulation are collected in Table~\ref{tab:evoTimes}. Although there is a small degree of stochasticity involved in the simulations owing to the use of radiative transfer methods, the effect is small (much less than one per cent on the total simulation duration) and the difference in the evolutionary times between the various models is indeed due to assumptions on the ice thickness and how it affects the dust and -- by extension -- gas temperatures. We note here again that in the present work, the gradual increase of the ice thickness is not simulated in the chemical model (i.e., the grain size remains fixed), as the associated increase in the grain surface area is small\footnote{The adsorption rates are proportional to the grain surface area, and so an increased surface area translates to a larger adsorption rate coefficient and hence faster depletion. Adding 80 monolayers of ice on a grain of radius 0.1\,$\mu$m increases the cross section by a factor of 1.5 only (assuming a 3 nm layer thickness).} and we do not expect large chemical effects in the present case with 0.1\,$\mu$m grains \citep[see also][]{Acharyya11}. Here we concentrate purely on the magnitude of the effect arising from opacity changes. In what follows, results from simulation~M\_80 are omitted for clarity of presentation; they are presented and discussed separately in Appendix~\ref{appendixA}.

The temporal evolution of $T_{\rm dust}$ is different in each simulation. To quantify this, Fig.\,\ref{fig:physEvoManyCells} shows the evolution of $T_{\rm dust}$ as a function of volume density in three selected model cells in simulations M\_TD, M\_10, and M\_0. The volume density increases with time in each plotted cell due to infall, and hence higher values of the volume density indicate later evolutionary times. In the innermost cell, which reaches a distance of 20\,au from the origin when $n({\rm H_2}) = 10^7 \, \rm cm^{-3}$, $T_{\rm dust}$ is initially lower in the M\_TD simulation compared to the other two, but the situation is reversed as the simulation progresses. This result is seemingly counterintuitive as one might expect the ice growth to lead to a lower value of $T_{\rm dust}$, which does indeed occur at lower volume densities. However, the increase of the ice thickness also leads to an increase of the optical depth, making it more difficult for the cooling radiation to escape the cloud. At the very highest densities, the differences between the temperature curves tend toward zero, indicating that the optical depth is so high that the ice thickness no longer has a notable contribution to $T_{\rm dust}$. The overall differences between the $T_{\rm dust}$ curves from the three simulations are small, however, as evidenced by the lower panel on the left-hand side of the Figure.

The transient spikes in $T_{\rm dust}$ are radiative transfer noise. The central region of the core is very small compared to its outer radius, and consequently the determination of $T_{\rm dust}$ is sometimes difficult due to an insufficient number of photon packets reaching the center. We have confimed via testing that the noise can be reduced by increasing the number of photon packets injected into the cloud, but at an added computational cost which will be significant over the course of the full simulation. A sudden increase in $T_{\rm dust}$ is of course expected to affect the grain-surface chemistry, and indeed we do see such effects for some species in our data (see Fig.\,\ref{fig:abundances_time}). The overall evolution of the cloud is however dictated mainly by regions outside the very center, and hence we consider the temperature noise to be acceptable for the purposes of the present simulations.

The temperature noise is at a much lower level in the cells that start at larger distances from the core center, as also displayed in Fig.\,\ref{fig:physEvoManyCells}, showing that it is indeed only the very center of the model cloud that is greatly affected by the accuracy of the radiative transfer simulations. Similar tendencies to the innermost cell are evident in the outer ones too; the dust is cooler in model~M\_TD for volume densities lower than approximately $10^5\,\rm cm^{-3}$, and the difference in $T_{\rm dust}$ between the non-dynamic ice models to M\_TD increases with decreasing volume density. In the cell ultimately reaching 1790\,au, there is a crossing of the temperature curves just like in the innermost cell, but the value of volume density for which the crossing occurs depends on the simulations being compared. The dust is always colder in model~M\_TD at low volume densities as shown in Fig.\,\ref{fig:physEvoManyCells} for the cell ultimately reaching 9290\,au.

To complement Fig.\,\ref{fig:physEvoManyCells}, we show in Fig.\,\ref{fig:physEvoTime} the radial $T_{\rm dust}$ profiles in the three simulations for two central volume densities: $10^5 \, \rm cm^{-3}$ and $10^7 \, \rm cm^{-3}$. The profiles are snapshots taken at the evolutionary time when the central density hits the given value in each simulation; the corresponding density profiles are also displayed for reference. These plots reinforce the conclusions made above based on Fig.\,\ref{fig:physEvoManyCells}. The dust temperature can be either lower or higher in the fully dynamic simulation~M\_TD as compared to simulations M\_10 and M\_0 -- typically, the dust is colder almost throughout the examined region in simulation~M\_TD as compared to the other two, but a reversal occurs near the center for high central densities, in accordance with the evolutionary curves shown in Fig.\,\ref{fig:physEvoManyCells}. Overall, the differences between the simulation results are quite small, however, and the $T_{\rm dust}$ profiles differ by a maximum of $\sim$0.5\,K only. Here we show the profiles up to a radius of 25,000\,au, that is, a quarter of the total radius of the core; at still higher radii, the $T_{\rm dust}$ profiles approach each other and by $\sim$40,000\,au the temperature differences have diminished to the 0.1\,K level. We note that $T_{\rm dust}$ increases slightly in the outer core in all simulations as the core collapses, which is a result of the decrease in volume density there as the core becomes more and more centrally concentrated, leading to an increase of the heating of the dust by the interstellar radiation field.

\begin{figure*}
\centering
        \includegraphics[width=2.0\columnwidth]{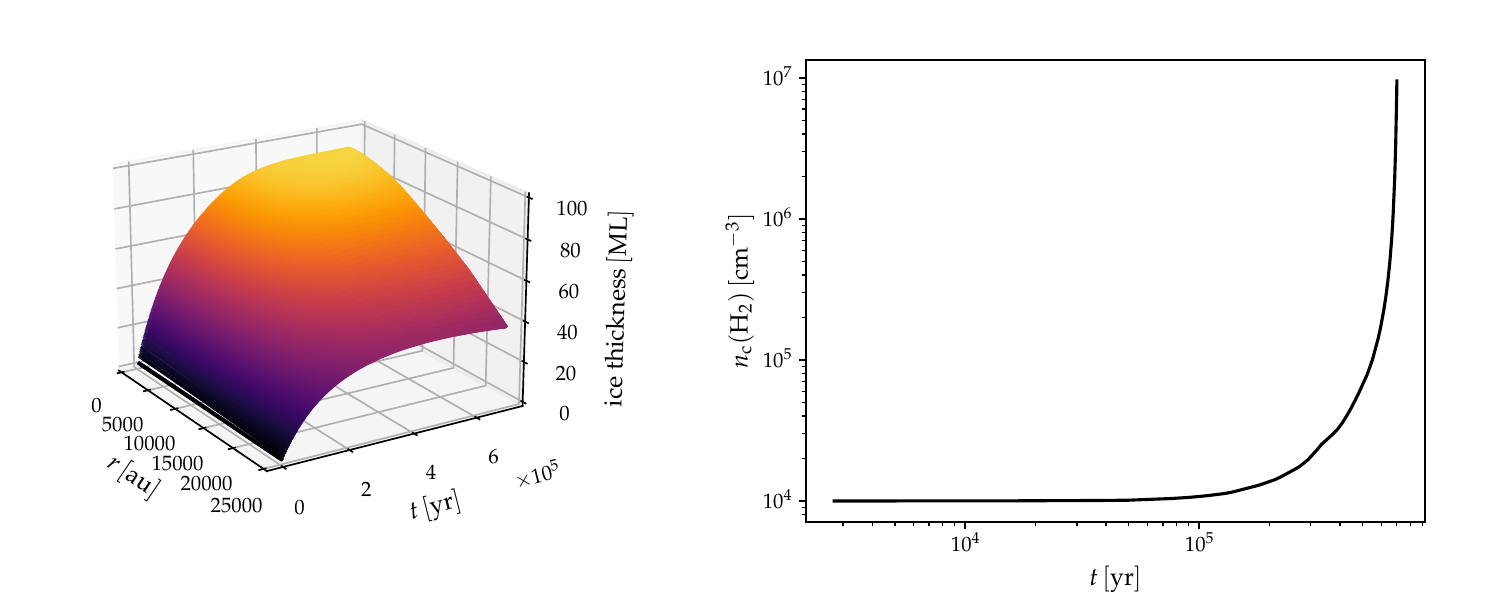}
    \caption{Left: thickness of the ice in ML in simulation~M\_TD as a function of radius and time. The colors are a guide to the eye; the maximum value of the ice thickness ($\sim$90 ML) is represented by yellow. Right: time evolution of the volume density at the center of the core, denoted by $n_{\rm c}(\rm H_2)$.
    }
        \label{fig:iceThickness}
\end{figure*}

The evolution of the ice thickness (in monolayers; ML) in simulation~M\_TD is shown in Fig.\,\ref{fig:iceThickness}. The thickness saturates at high volume densities because a balance between adsorption and (non-thermal) desorption is reached, and the saturation value of approximately 90~monolayers is reached already well before the central density hits $n({\rm H_2}) = 10^5 \, \rm cm^{-3}$ (the saturation value depends on the chosen values of the grain parameters). The data show clearly that the ice thickness increases to several tens of monolayers already very early into the evolution of the core: at a central density of only $n({\rm H_2}) = 2 \times 10^4 \, \rm cm^{-3}$, or a simulation time of $t \sim 3 \times 10^5\,\rm yr$, the thickness is already approximately 80 monolayers in the core center, and approximately 40 monolayers at 25,000 au (or $n({\rm H_2}) \sim 4 \times 10^3\, \rm cm^{-3}$; cf. Fig.\,\ref{fig:physEvoTime}). This suggests that dust models corresponding to thin ices (on the order of 10~ML) are not appropriate in molecular cloud conditions except at the very earliest evolutionary times.

In the present model that does not include any magnetic field or turbulence effects, the core is supported against gravitational collapse by the thermal pressure. Given that there are differences in the collapse timescale in the various simulations (Table~\ref{tab:evoTimes}), there must be variations in the gas temperature among the simulations as well. Indeed we find this to be the case; the magnitude of the maximum gas temperature variations is similar -- approximately on the order of 0.5 \,K -- to the one shown for $T_{\rm dust}$ in Fig.\,\ref{fig:physEvoTime} going from simulation~M\_TD to M\_0, with the gas temperature being the lowest in simulation~M\_TD. Examination of the simulation data shows that the gas temperature differences are caused by the dust opacity variations and their effect on the photoelectric heating efficiency; the effect of $T_{\rm dust}$ variations on the abundances of coolant molecules (see Sect.\,\ref{ss:molecularAbundances}) is very small in comparison. We omit the gas temperature plots here for brevity, but the effect of photoelectric heating is quantified in Sect.\,\ref{s:discussion} and in Appendix~\ref{appendixA}. Given that the core is thermally supported, the fact that the M\_TD simulation presents the lowest average gas temperature makes sense in the context of the collapse times (Table~\ref{tab:evoTimes}), as a lower temperature means less thermal support and hence faster collapse. It must however be emphasized that the differences in the collapse time among the simulations is quite insignificant (the exception being simulation~M\_80; see Appendix~\ref{appendixA}).

\subsection{Molecular abundances in the gas phase and in the ice}\label{ss:molecularAbundances}

\begin{figure*}
\centering
        \includegraphics[width=2.0\columnwidth]{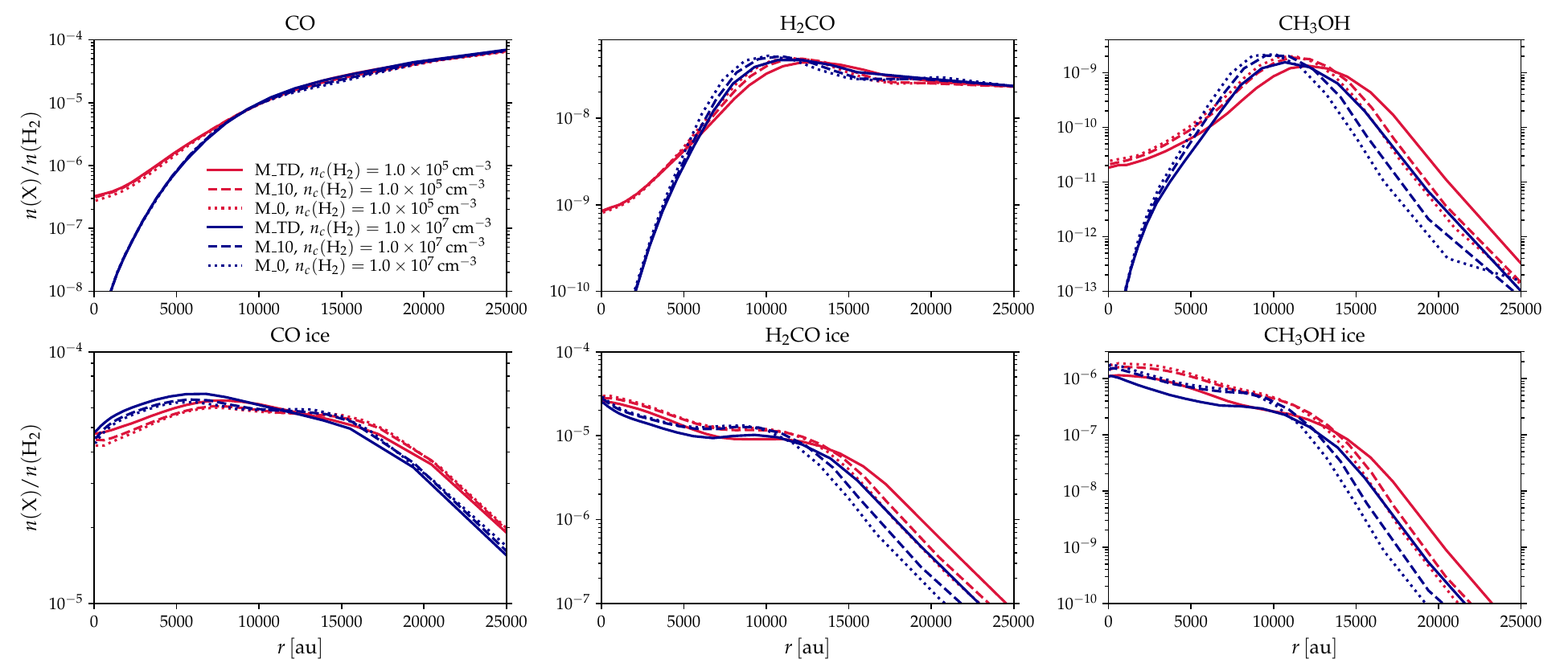}
    \caption{Abundances of selected gas-phase (top panels) and ice (bottom panels) molecules with respect to $\rm H_2$ as a function of radius in simulations~M\_TD, M\_10 and M\_0. Profiles are shown for two different values of the central $\rm H_2$ density: $10^5$ (red) and $10^7\,\rm cm^{-3}$ (blue). Solid, dashed, and dotted lines correspond to simulation~M\_TD, M\_10, and M\_0, respectively. The central density of $10^5\,\rm cm^{-3}$ corresponds to $t = 5.50\times10^5\,\rm yr$, $t = 5.57\times10^5\,\rm yr$, and $t = 5.74\times10^5\,\rm yr$ in simulations~M\_TD, M\_10 and M\_0, respectively. The corresponding values for central density $10^7\,\rm cm^{-3}$ are given in Table~\ref{tab:evoTimes}.}
        \label{fig:abundances}
\end{figure*}

\begin{figure*}
\centering
        \includegraphics[width=2.0\columnwidth]{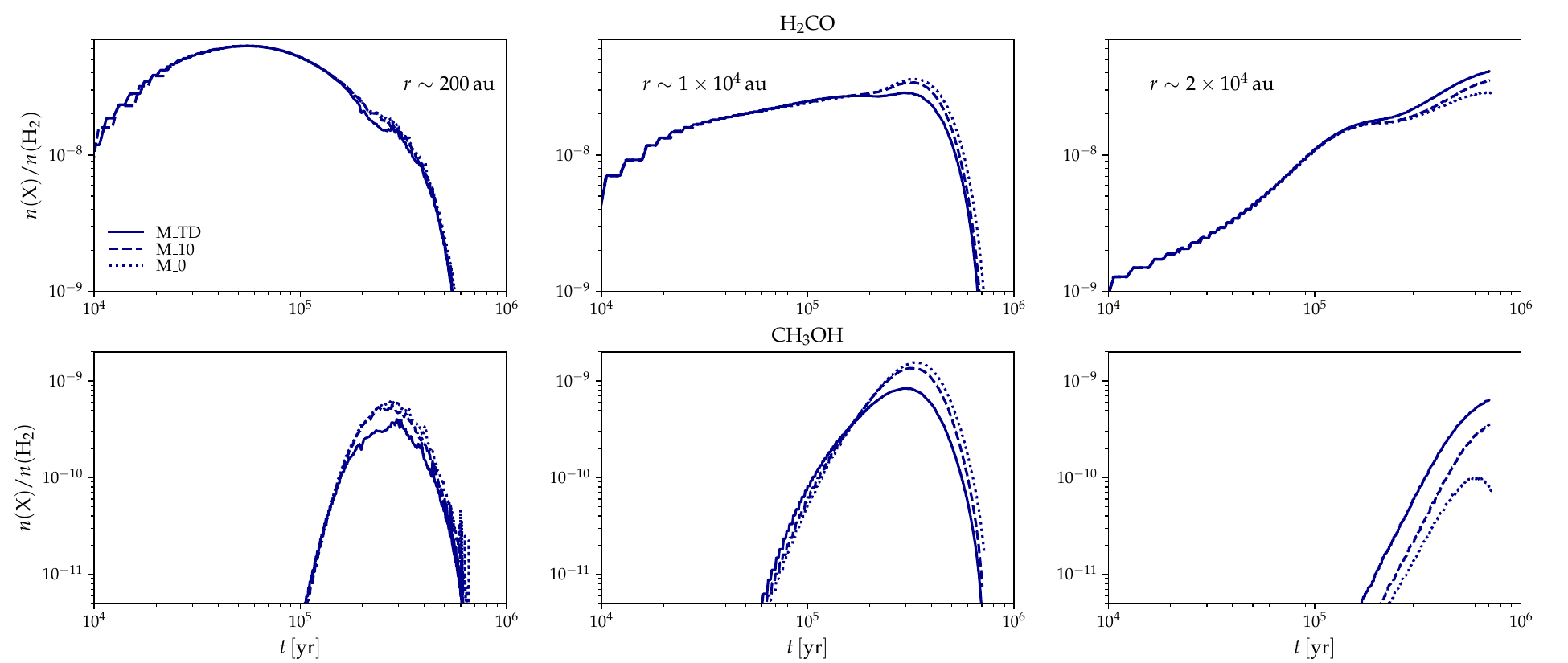}
    \caption{Abundances of gas-phase $\rm H_2CO$ (top panels) and $\rm CH_3OH$ (bottom panels) with respect to $\rm H_2$ as a function of time in simulations~M\_TD (solid lines), M\_10 (dashed lines) and M\_0 (dotted lines). From left to right, the columns show the evolution in a cell that starts at $r \sim 200\,\rm au$, $r \sim 1\times10^4\,\rm au$, or $r \sim 2\times10^4\,\rm au$. The noise in the $\rm CH_3OH$ abundances in the innermost cell reflects the temperature fluctuations (Fig.\,\ref{fig:physEvoManyCells}), which affect the abundance of hydrogen on the grain surface (gas-phase $\rm CH_3OH$ is mostly coming from chemical desorption of methanol formed by the $\rm CH_3O + H$ reaction on the grains).}
        \label{fig:abundances_time}
\end{figure*}

It is clear from the above that the differences in $T_{\rm dust}$ between the various simulations are small and rarely exceed 0.5\,K. The efficiency of chemical reactions on grain surfaces, especially ones involving direct hydrogenation, is determined by a competition between diffusion and desorption rates. Given the inverse exponential dependence of the diffusion rates of atoms and molecules on $T_{\rm dust}$, even such small temperature variations may have an effect on molecular abundances on grain surfaces, and by extension possibly in the gas phase via desorption, so long as the reactants are available. A priori, one may expect largest effects for molecules whose formation depends on barrier-mediated reactions as the residence time of the reactants is a critical factor.

Figure~\ref{fig:abundances} shows the abundances of three molecules along the main methanol formation path: CO, $\rm H_2CO$, and $\rm CH_3OH$ itself. Simulations~M\_TD~and~M\_10 predict similar abundances for all of the plotted species. At 20,000 au, the difference in methanol abundance is approximately a factor of three, though in this region the abundance is too low to be of observational consequence. At the location of the peak methanol abundance, the difference is only few tens of per cent. Ice abundances show somewhat larger differences, but they too are only on the factor of two level in the inner core. Figure~\ref{fig:abundances} confirms the expectation that abundance variations are larger for species that depend on barrier-mediated reactions to form (under the assumption that surface chemistry is governed by diffusive reactions), with CO and $\rm H_2CO$ showing quite similar gas-phase abundances in the two simulations\footnote{$\rm H_2CO$ formation from H + HCO also involves a barrier, but it is much smaller than the one for the step leading to methanol.}. Extending the comparison to simulation~M\_0 reveals similar trends, just with somewhat larger abundances differences, for example a factor of about five (as opposed to three) at 20,000 au, using simulation~M\_TD as the baseline.

Another trend that is evident in Fig.\,\ref{fig:abundances} is that the gas-phase $\rm H_2CO$ and methanol abundances are higher in simulations~M\_10~and~M\_0 than in M\_TD in the inner core (5,000 to 10,000\,au), but lower in outer regions, despite the fact that in simulation~M\_TD, $T_{\rm dust}$ is lower throughout (see Fig.\,\ref{fig:physEvoTime}) -- gas-phase methanol is produced mostly via chemical desorption from the grains and hence the trends in the gas-phase population correlate with those in the ice. The lower $T_{\rm dust}$ reduces the mobility of hydrogen atoms on the surface with the effect that, in the inner core, methanol production in the ice is less efficient in simulation~M\_TD compared to the other two. The trend is reversed at larger radii where the lower $T_{\rm dust}$ helps to maintain hydrogen atoms on the surface long enough to form methanol. Still, the differences between the abundances predicted by the three simulations are only a factor of 2-3 at most, diminishing in the outer core (for radii larger than $\sim$40,000\,au the temperature is high enough even in simulation~M\_TD that the abundance profiles approach each other), and beyond $\sim$15,000\,au the methanol abundance is too low to be observationally significant anyway. We note that the gas-phase methanol abundance peak occurs at $\sim$10,000\,au, or a volume density of $\sim 2 \times 10^4 \, \rm cm^{-3}$, regardless of the simulation, which is compatible with observations toward cold cores \citep[e.g.,][]{Bizzocchi14,Spezzano16b, Harju20}. This observational finding has also been corroborated via simulations by \citet{Vasyunin17} and \citet{Riedel23}.

To complement Fig.\,\ref{fig:abundances}, we show in Fig.\,\ref{fig:abundances_time} the time evolution of $\rm H_2CO$ and $\rm CH_3OH$ in three cells in simulations M\_TD, M\_10~and~M\_0. The same trends as in Fig.\,\ref{fig:abundances} are present here as well, for example, that methanol is the least abundant in simulation~M\_TD at 10,000\,au, but the situation is reversed at larger radii. Figure~\ref{fig:abundances_time} also shows that differences in abundances between the simulations generally start to appear after $t \sim 10^5 \, \rm yr$. Also, we see clearly the effect of the ice thickness; in the inner core where the ice is thick already at early times in simulation~M\_TD (Fig.\,\ref{fig:iceThickness}), the abundances predicted by simulation~M\_TD differ by a factor of 2-3 as compared to the other two simulations, but the difference between the solutions diminishes at larger radii where the ice thickness in simulation~M\_TD (a few tens of ML; Fig.\,\ref{fig:iceThickness}), is closer to the thickness in simulations~M\_10~and~M\_0. Although the overall differences are on the factor of a few level only, the fully time dependent simulation~M\_TD does predict distinctly different abundances compared to simulations where the ice thickness does not change with time.

Figure~\,\ref{fig:abundances} shows that even though the abundance variations from simulation to simulation for the various ice species are small, the abundances do vary in different directions (e.g., CO abundance decreases while methanol abundance increases as the ice mantle is made thinner), with potential observational implications. We discuss the ice populations in more detail in Sect.\,\ref{s:discussion}. Finally, we note that the time-dependence of the effect of $T_{\rm dust}$ variations is also quite weak, as seen by comparing the red and blue curves in Fig.\,\ref{fig:abundances} -- though the comparison is not straightforward because the abundance curves at different times are affected also by the gas parcels drifting inward, as evidenced by the gas-phase $\rm H_2CO$ and methanol abundance peaks shifting with time.

%Show still: evolution of extinction curves?

\section{Discussion}\label{s:discussion}

\begin{figure}
\centering
        \includegraphics[width=0.8\columnwidth]{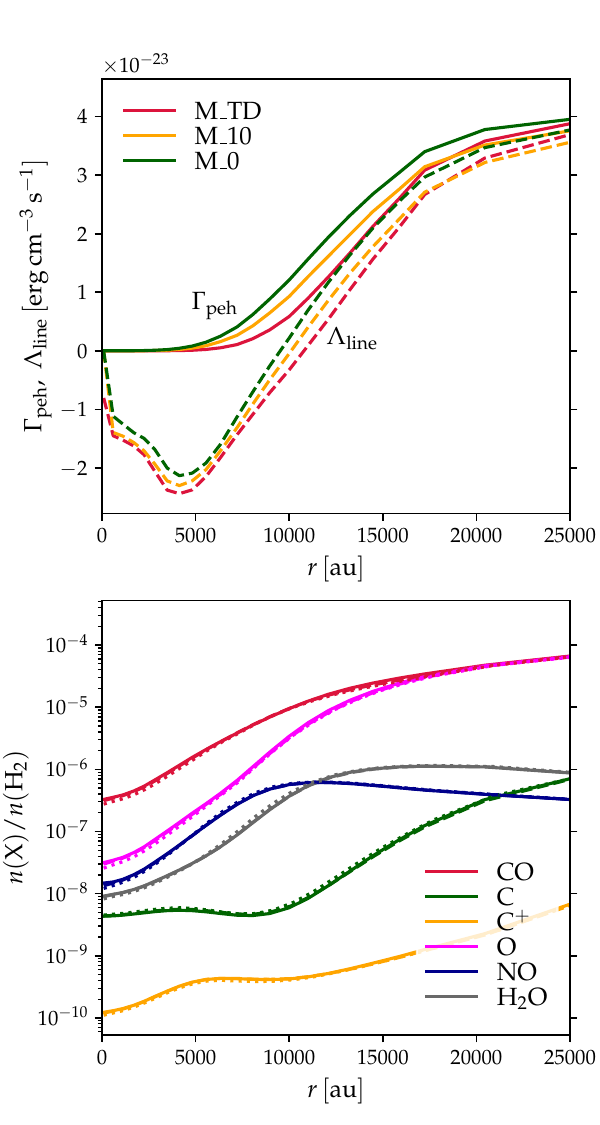}
    \caption{Line cooling ($\Lambda_{\rm line}$; dashed lines) and photoelectric heating ($\Gamma_{\rm peh}$; solid lines) rates as a function of radius at the time when the central density is $n({\rm H_2}) = 10^5 \, \rm cm^{-3}$ in simulations M\_TD, M\_10, and M\_0 (top panel). Radial distributions of selected cooling molecules at the same time (bottom panel). In this panel, solid, dashed, and dotted lines represent simulations~M\_TD, M\_10, and M\_0, respectively, but they overlap almost perfectly.
    }
        \label{fig:coolHeat}
\end{figure}

Our simulations confirm that the time-dependent growth of ices on dust grains does have an effect on the gravitational collapse of a molecular cloud core, but that the effect is quite limited. An effect arises because changing the dust opacity influences the efficiency of photoelectric heating, which in turn modifies the gas temperature and hence the thermal pressure at large scales. Figure~\ref{fig:coolHeat} shows the photoelectric heating and line cooling rates in the three simulations at the time when the central density is $n({\rm H_2}) = 10^5 \, \rm cm^{-3}$. The heating rate tends to decrease with increasing ice thickness, though the trend is not linear with ice thickness and at low volume densities the curve from simulation~M\_TD crosses that of simulation~M\_10 even though the ice thickness is above 40\,ML at these radii (Fig.\,\ref{fig:iceThickness}). The Figure shows that the line cooling efficiency also changes from simulation to simulation, but this is not due to changes in the abundances of the main coolants as evidenced by the bottom panel. We show in the plot a subset of all coolant molecules (see \citealt{Sipila18} for the full list), but we have checked that similarly negligible abundance variations occur for the rest of the coolants as well. Instead, the variations in the line cooling are a response to the gas temperature changes brought about by the photoelectric heating. The strength of the other cooling and heating mechanisms in the model, namely heating by cosmic rays and cooling by the gas-dust collisional coupling, does not vary much from simulation to simulation because they are tied to the volume density profile. The photoelectric heating rate variations are well within a factor of two, which explains why the physical evolution of the cores, as quantified here by the time to reach a central density of $n({\rm H_2}) = 10^7 \, \rm cm^{-3}$ (Table~\ref{tab:evoTimes}), is so similar in all cases. The line cooling rate obtains negative values in the central core. This is because of a large optical depth which makes it difficult for the line radiation to escape, causing it to heat the gas instead of cooling it; see \citet{Sipila18} for further discussion of the heating and cooling and a breakdown of the typical line cooling rates of the coolants included in the model.

\begin{figure*}
\centering
        \includegraphics[width=2.0\columnwidth]{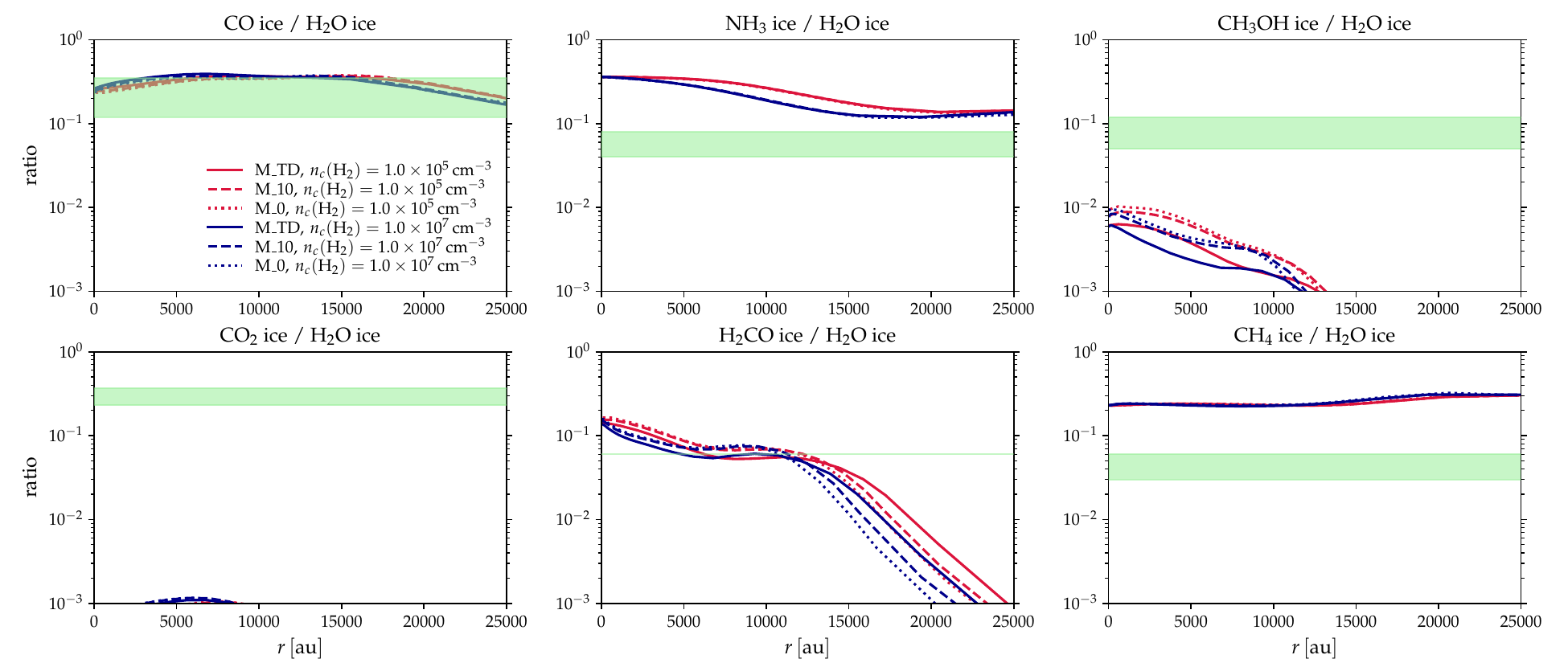}
    \caption{Abundance ratios of selected ice molecules to water ice as a function of radius in simulations~M\_TD, M\_10 and M\_0, plotted on a common y-axis scale to ease the comparison. Profiles are shown for two different values of the central $\rm H_2$ density: $10^5$ (red) and $10^7\,\rm cm^{-3}$ (blue). Solid, dashed, and dotted lines correspond to simulation~M\_TD, M\_10, and M\_0, respectively. The light green bands show the ratios derived from observations toward low-mass young stellar objects \citep{Boogert15}.}
        \label{fig:abundances_iceratios}
\end{figure*}

Figure~\ref{fig:abundances} hints at changes in the relative populations of ices when the dust opacity model is varied. We show in Fig.\,\ref{fig:abundances_iceratios} ice abundance ratios (with respect to water) as predicted by the three simulations. We concentrate here on selection of species that have been observed either earlier \citep{Boogert15} or more recently with the JWST \citep{McClure23, Rocha24}. First, it is clear that once again the variations in the ratios from simulation to simulation are small, within a factor of two at most (typically smaller than the observational errors) and that variations only occur for species that are formed via barrier-mediated reactions in the ice. Second, the predictions of our simulations are in partial disagreement with observed ice abundance ratios, as only the observed $\rm CO/H_2O$ and $\rm H_2CO/H_2O$ ratios are reproduced. The simulations tend to overproduce molecules that form via barrierless hydrogenation reactions, while underproducing molecules that require barrier-mediated reactions to form. The $\rm CO_2/H_2O$ ratio is underestimated by over two orders of magnitude. There are several possible reasons for the discrepancies, and we cite here a few possibilities. We are presently using a ``two-phase'' model where the ice is treated as a single reactive layer, while ``three-phase'' models that separate the ice into a reactive surface layer and a (possibly inert) bulk beneath will very likely give different values for the various abundance ratios \citep{Hasegawa93b}. The diffusion-to-binding energy ratio ($E_{\rm d}/E_{\rm b}$) governs the mobility of the surface species in a diffusion-based model like ours; here we use a value of 0.6 while there is a large range of values explored in the literature (0.2 to 0.8; the value may also be species-dependent, see \citealt{Furuya22}). The treatment of chemical desorption will affect the ice (and gas-phase) populations \citep{Vasyunin17, Riedel23}. Nondiffusive chemistry is expected to boost complex molecule formation and this should occur at the expense of reactions that proceed via barrierless hydrogenation. Finally, the assumed elemental abundances affect the simulation results; for example the high simulated ammonia abundance may be due in part to a higher elemental nitrogen abundance compared to the observed regions. The dust temperature plays a direct or indirect role in most of these processes, but nevertheless based on the present results we have little reason to expect that our general conclusions would be altered significantly even if the simulation parameter space was varied beyond the variation of the dust opacity -- keeping in mind that with the present model we are only able to predict the abundances of molecules up to methanol in complexity. It is still conceivable that larger abundance differences between the simulations could be found for more complex molecules whose formation depends on more elaborate barrier-mediated reaction networks. As an example of the possible parameter space variations, one may for example consider a different dust grain material or a mix of them as was done here, or different refractive index data for the ice itself. One could also consider a mix of ice materials, perhaps evolving with time as the ice composition changes in the chemical model (this exercise is however severely limited by the availability of refractive index data for ices). Such changes would alter the equilibrium dust temperature and hence the abundances of the various molecules in the ice, but the effect on the collapse of the core with respect to the present simulations would probably be very limited, although quantitative claims cannot be made without performing additional simulations.

In this paper we have made the fundamental assumption that all grains have the same size, that is, a radius of 0.1\,$\mu$m. In reality one expects instead a distribution of grain sizes, and indeed various size distributions have been invoked to explain observations of dust emission \citep[e.g.,][]{Mathis77,Weingartner01,Jones13}. We have shown in a previous work \citep{Sipila20} that: 1) when switching from a monodisperse grain model to a distribution, the ice composition depends on the grain size in a non-linear fashion; 2) the various grain populations have a different equilibrium temperature and, due to extinction effects, the smallest grains can be the warmest at low densities, but the coolest in inner regions; and 3) the ice thickness varies by a factor of $\sim$two over the distribution. All of these factors taken together may lead to complex effects when considering the entire evolution of a cloud core as is done in the present paper (in \citealt{Sipila20} we used a static cloud model). Investigating the effect of a grain size distribution in the context of a collapsing core, with time-dependent dust opacity, will the subject of a follow-up work.

\section{Conclusions}\label{s:conclusions}

We investigated the effect of time-dependent ice growth on the dust opacity and hence on the dust temperature in a collapsing cloud core. To accomplish this, we employed our hydrodynamical code HDCRT that self-consistently combines hydrodynamics, chemistry, and radiative transfer simulations. We assumed for the sake of simplicity spherical grains with a fixed grain material composition (carbon/silicate core in a 40/60 mass ratio) and the opacity calculations were carried out assuming that the ice is made up of water only. We ran four simulations, three of which assumed a time-independent dust model corresponding to 0, 10, or 80 monolayers of ice on the grains, while in the fourth simulation the opacity was allowed to vary with time according to the ice growth. The collapse was simulated in a one-dimensional framework assuming spherical symmetry.

The dust temperature affects the abundances of molecules in the ices on dust grains because chemical reactions are rate-limited by the diffusion of the reactants. We find that the variations in the dust temperature profiles brought about by the different assumptions on ice opacity have a small effect on the abundances of species that are formed via barrier-mediated chemical creations, such as methanol, owing to changes in the competition between diffusion and desorption rates. The gas-phase populations present variations on a similar order of magnitude due to (chemical) desorption of the ice molecules. However, the abundances of molecules that are formed via barrierless reactions are very similar in all simulations regardless of the assumed ice thickness.

The dust opacity variations also affect the collapse timescale because of the induced changes in photoelectric heating efficiency, but this effect is also small. Here we tracked the time it takes for the core to reach a central $\rm H_2$ density of $10^7 \, \rm cm^{-3}$ -- the difference in the time to reach this density between the slowest and fastest collapsing solutions is only $\sim$15 per cent. The collapse dynamics are dominated by the large-scale structure and hence the simulation assuming 80 monolayers of ice proceeds the fastest, because the dust is relatively cold in the outer core as compared to the other simulations.

The results of the fully time-dependent simulation, with dust opacity changing as a function of ice thickness, demonstrate that the ice thickness increases quite rapidly; in the center of the core, a thickness of 70~monolayers is reached in a time of just over $2 \times 10^5 \, \rm yr$ when the central density of the core is only $1.5 \times 10^4 \, \rm cm^{-3}$ (the initial value in the simulation is $1.0 \times 10^4 \, \rm cm^{-3}$). The thickness saturates to a value of $\sim$90 monolayers as a result of a balance between adsorption and desorption. (The saturation value depends on the assumed dust parameters.) In the outer core, the ice is several tens of monolayers thick out to a distance of $\sim$30,000\,au from the center, and only at a density of $\sim$$10^3 \, \rm cm^{-3}$ ($\sim$40,000\,au from the center) does it decrease below 10 monolayers. Therefore, our results clearly preclude the use of thick (above 50) or thin (approximately ten monolayers) ices throughout when simulating an interstellar cloud core. \citet{Hocuk17} for example have reached similar conclusions. While these assumptions do not have a significant impact on the dust temperature and on the chemistry, as demonstrated here, this conclusion could be of importance in simulations of dust emission.

We have found in a previous work \citep{Sipila20} that molecular abundances in the ice depend non-linearly on the grain size when a size distribution is introduced in a chemical model, as opposed to the assumption of constant size as is customarily done, also in the present paper. The ice thickness will also obtain different values depending on the grain size. In a follow-up paper, we will extend the present analysis to size distributions to quantify the magnitude of the effect of time-dependent ice growth in a more realistic scenario involving multiple grain sizes, also including in the chemical model the temporal increase in grain size brought about by ice growth (such effects have been previously included in models by, e.g., \citealt{Acharyya11,Kalvans15c}). That analysis will also cover a larger set of ice molecules than the present model, with which we could reliably simulate the evolution of molecules up to methanol in complexity.

%thin ice mantles inappropriate, connect with dust emission

\begin{acknowledgements}
We thank the anonymous referee for constructive comments and suggestions. O.S. and P.C. gratefully acknowledge the funding by the Max Planck Society. M.J. acknowledges support from the Research Council of Finland grant No. 348342.
\end{acknowledgements}

\bibliographystyle{aa}
\bibliography{iceGrowth_I_accepted.bib}

\appendix

\onecolumn

\section{Results of the M\_80 simulation}\label{appendixA}

\begin{figure*}
\centering
        \includegraphics[width=0.9\columnwidth]{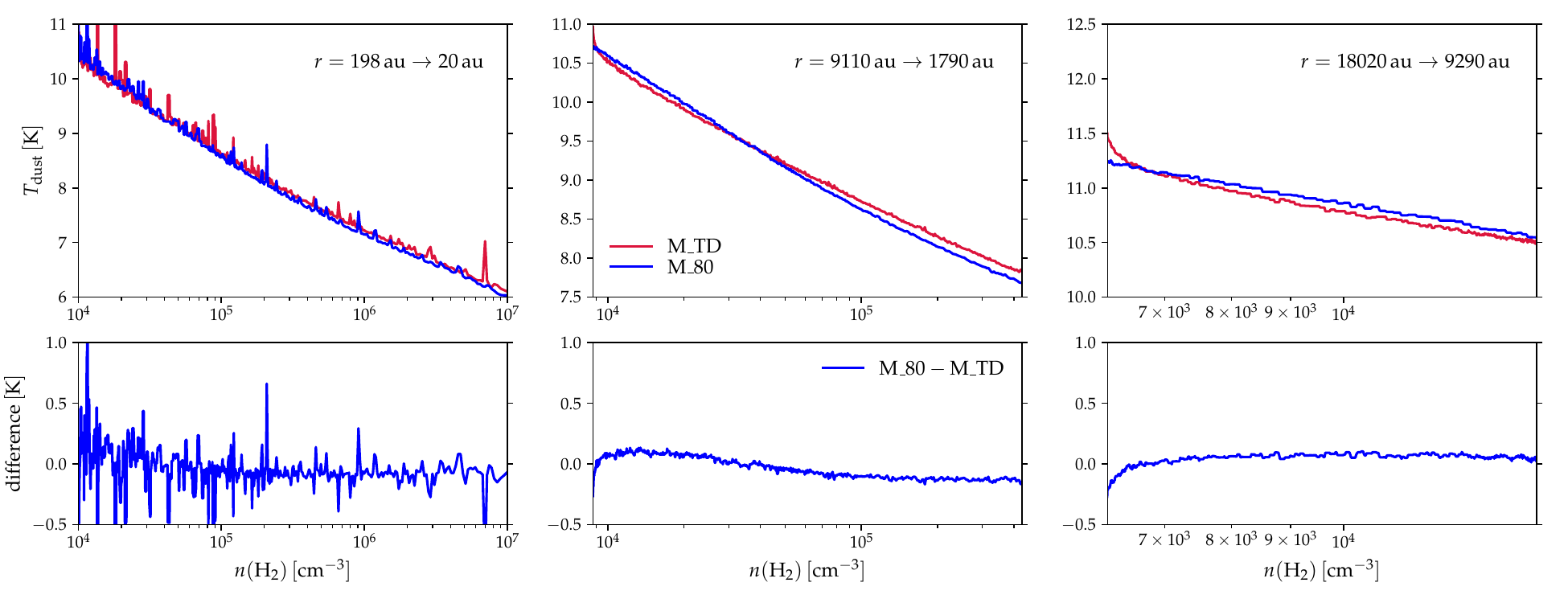}
    \caption{As Fig.\,\ref{fig:physEvoManyCells}, but for simulations~M\_TD~and~M\_80.}
        \label{fig:physEvoManyCells_80ML}
\end{figure*}

\begin{figure}
\centering
        \includegraphics[width=0.35\columnwidth]{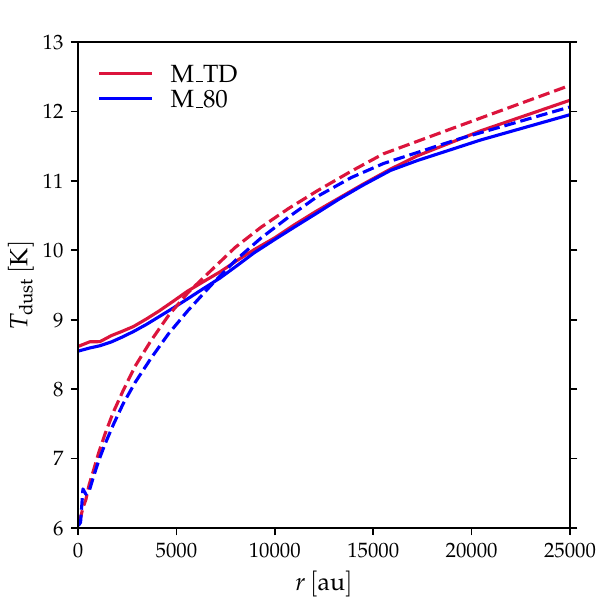}
    \caption{Snapshots of the dust temperature as a function of radius in simulations~M\_TD and M\_80 at the time when the central density of the core hits either $n({\rm H_2}) = 10^5 \, \rm cm^{-3}$ (solid lines) or $n({\rm H_2}) = 10^7 \, \rm cm^{-3}$ (dashed lines). In contrast to Fig.\,\ref{fig:physEvoTime}, the differences between the curves are not shown separately as they are very minor (on the order of 0.1\,K).}
        \label{fig:dustTemperature_80ML}
\end{figure}

\begin{figure*}
\centering
        \includegraphics[width=0.9\columnwidth]{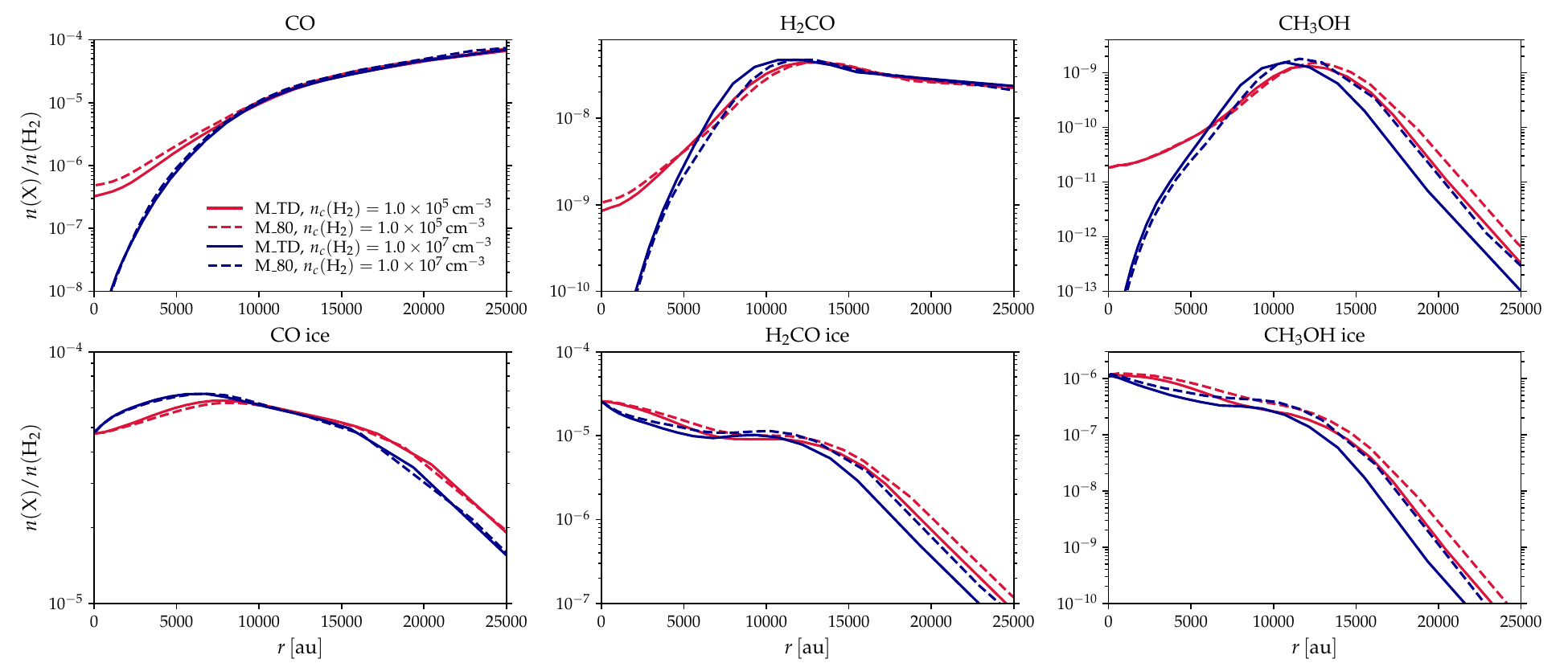}
    \caption{As Fig.\,\ref{fig:abundances}, but for simulations~M\_TD~and~M\_80.}
        \label{fig:abundances_80ML}
\end{figure*}

\begin{figure}
\centering
        \includegraphics[width=0.35\columnwidth]{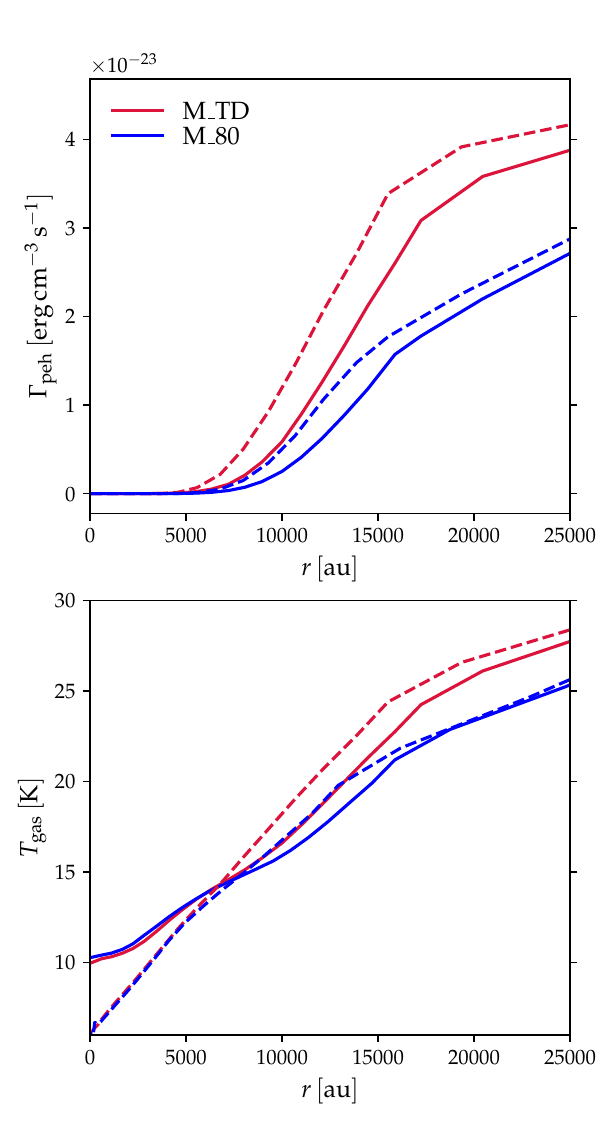}
    \caption{Snapshots of the photoelectric heating rate (top panel) and gas temperature (bottom panel) as functions of radius in simulations~M\_TD and M\_80 at the time when the central density of the core hits either $n({\rm H_2}) = 10^5 \, \rm cm^{-3}$ (solid lines) or $n({\rm H_2}) = 10^7 \, \rm cm^{-3}$ (dashed lines).}
        \label{fig:coolHeat_80ML}
\end{figure}

We show here the results of the M\_80 simulation to complement the discussion in the main text. Figures~\ref{fig:physEvoManyCells_80ML}~and~\ref{fig:dustTemperature_80ML} show the evolution of $T_{\rm dust}$ in selected cells as well as the snapshots of the radial profiles analogously to Figs.\,\ref{fig:physEvoManyCells}~and~\ref{fig:physEvoTime} in the main text. Figure~\ref{fig:abundances_80ML} shows the abundances analogously to Fig.\,\ref{fig:abundances}. It is evident that as far as the equilibrium $T_{\rm dust}$ and the molecular abundances are concerned, the results of this simulation are very close to those of M\_TD; the $T_{\rm dust}$ values predicted by the two simulations tend to be within 0.1\,K of each other depending on the time and location in the core. The closeness of the results is expected given how thick the ices are in simulation~M\_TD already at early evolutionary times.

Table~\ref{tab:evoTimes} shows that the M\_80 simulation proceeds on a somewhat shorter timescale compared to the other simulations. This is because of photoelectric heating; while the thicker ice mantle at radii approximately 10,000 au onward does not lead to significant differences in equilibrium $T_{\rm dust}$ between the two simulations, it does have a marked effect on the photoelectric heating rate as evidenced by Fig.\,\ref{fig:coolHeat_80ML}. The resultant lower gas temperature in simulation~M\_80 translates to less thermal support and a faster collapse. Given that even in this case the various molecular abundances do not vary significantly from one simulation to the other, we conclude that from the point of view of chemical simulations the choice of the dust opacity model is not a critical consideration. To obtain the most accurate representation of the collapse, one should consider the radial dependence of the ice thickness, however.

\section{Variations in the initial hydrogen abundance}\label{appendixB}

The core model used here extends to approximately 100,000\,au, where the volume density is $n({\rm H_2})\sim 2.4\times10^2 \,\rm cm^{-3}$ and the visual extinction is $A_{\rm V} = 1\,\rm mag$. In these conditions, our fiducial assumption of hydrogen being completely in molecular form (Table~\ref{tab:initialAbundances}) initially may not be appropriate. To explore the potential effect of changing the initial chemical conditions, we have run a variation of the M\_TD simulation where the initial hydrogen reservoir is distributed among atomic and molecular hydrogen in a 50:50 ratio, which means an initial abundance of 0.5 for atomic hydrogen, and 0.25 for molecular hydrogen. This model variant is denoted M\_TD(50:50).

\begin{figure*}
\centering
        \includegraphics[width=0.9\columnwidth]{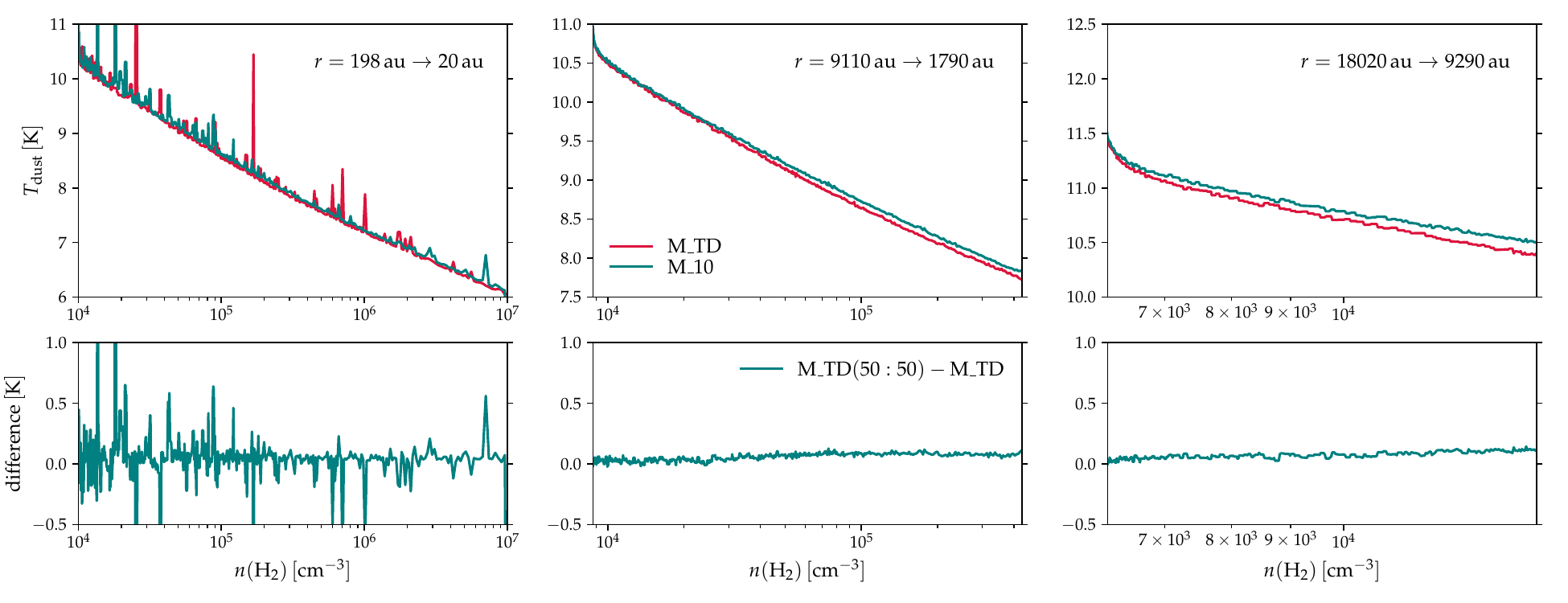}
    \caption{As Fig.\,\ref{fig:physEvoManyCells}, but comparing simulations~M\_TD~and~M\_TD(50:50).}
        \label{fig:physEvoManyCells_halfAtomicH}
\end{figure*}

\begin{figure}
\centering
        \includegraphics[width=0.35\columnwidth]{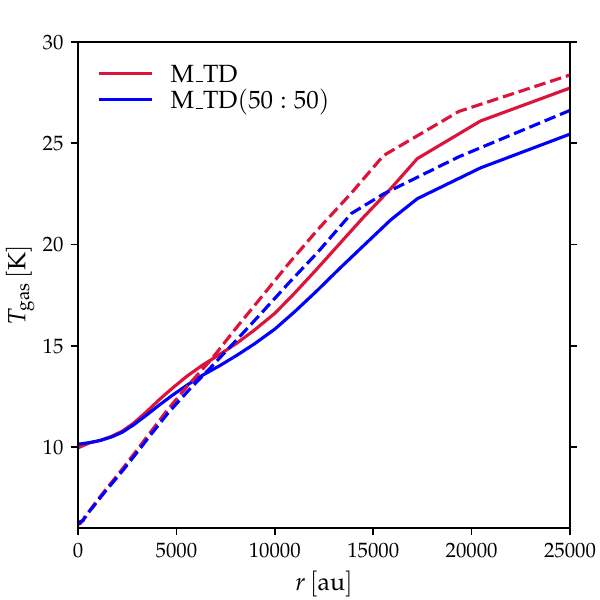}
    \caption{As the bottom panel in Fig.\,\ref{fig:coolHeat_80ML}, but comparing simulations~M\_TD~and~M\_TD(50:50).}
        \label{fig:gasTemperature_halfAtomicH}
\end{figure}

\begin{figure*}
\centering
        \includegraphics[width=0.9\columnwidth]{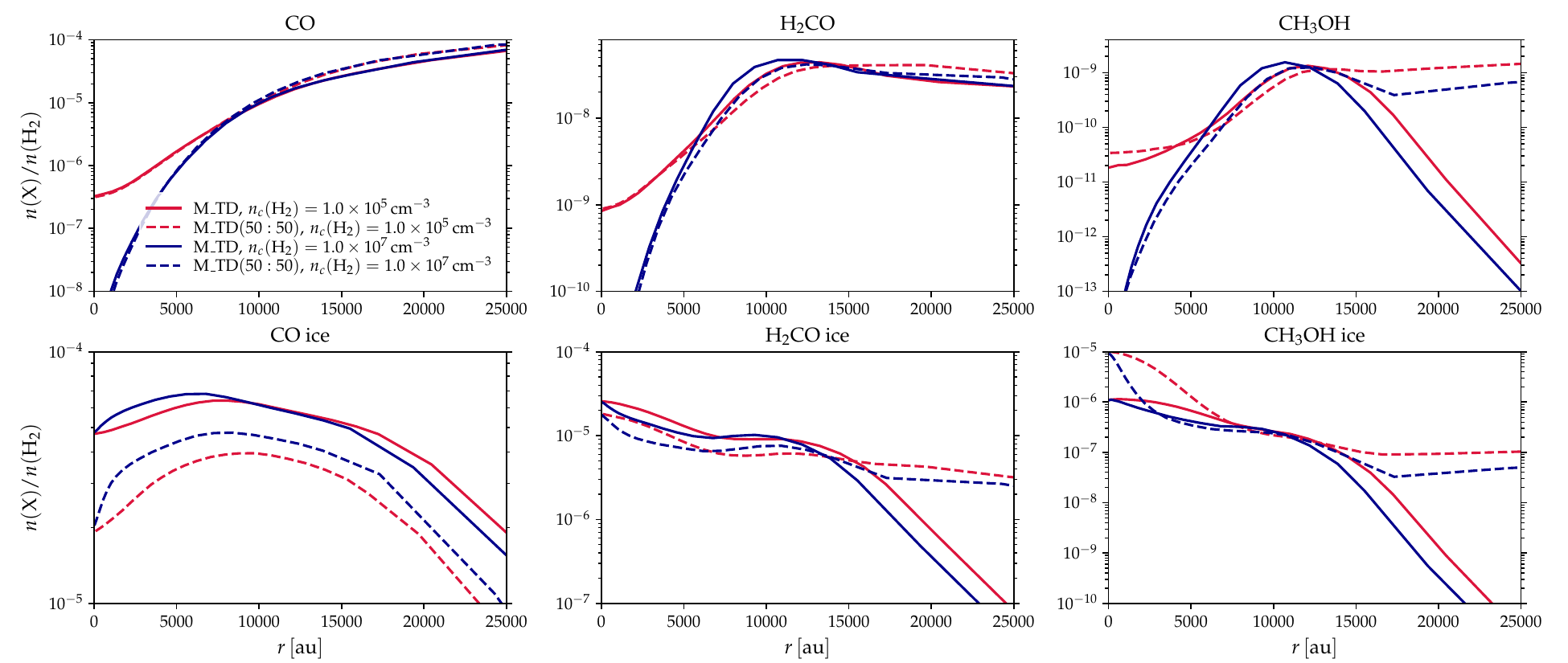}
    \caption{As Fig.\,\ref{fig:abundances}, but comparing simulations~M\_TD~and~M\_TD(50:50).}
        \label{fig:abundances_halfAtomicH}
\end{figure*}

\begin{figure*}
\centering
        \includegraphics[width=0.9\columnwidth]{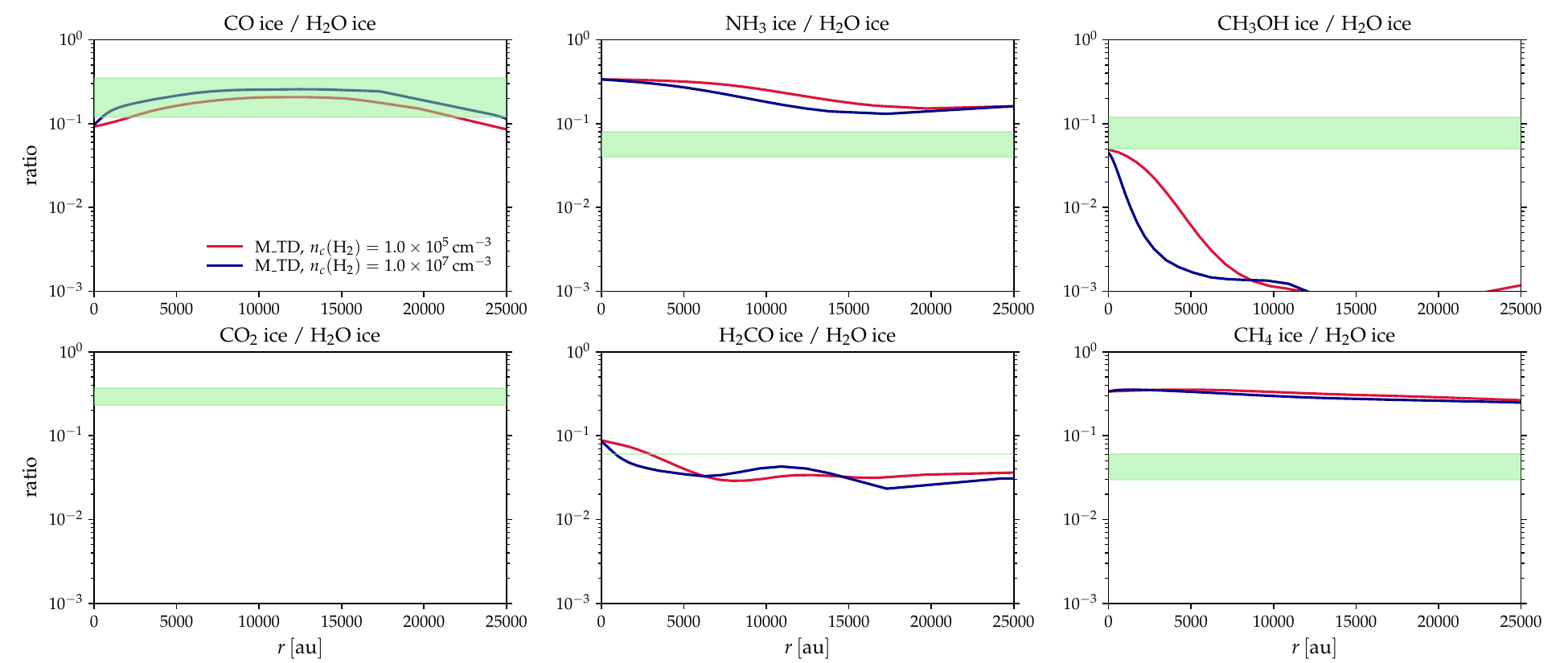}
    \caption{As Fig.\,\ref{fig:abundances_iceratios}, but showing only results from the M\_TD(50:50) simulation.}
        \label{fig:abundances_iceratios_halfAtomicH}
\end{figure*}

Figures~\ref{fig:physEvoManyCells_halfAtomicH} to \ref{fig:abundances_iceratios_halfAtomicH} show the results of simulation~M\_TD(50:50) in comparison to M\_TD. The initial chemical conditions, as far as the hydrogen content is concerned, have only a very minor effect on the physical evolution of the core. The central volume density reaches $10^7\,\rm cm^{-3}$ in $6.91 \times 10^5 \, \rm yr$ in simulation~M\_TD(50:50), as opposed to $7.00 \times 10^5 \, \rm yr$ in simulation~M\_TD. Figure~\ref{fig:physEvoManyCells_halfAtomicH} shows that there is a maximum $T_{\rm dust}$ variation of 0.1\,K between the M\_TD and M\_TD(50:50) simulations (disregarding the noise in the inner core). The gas temperature (Fig.\,\ref{fig:gasTemperature_halfAtomicH}) is somewhat lower in the latter simulation owing to an increased abundance of CO (Fig.\,\ref{fig:abundances_halfAtomicH})\footnote{CO is the main coolant in the plotted region, replaced by atomic C and finally $\rm C^+$ at increasing radii.}, which leads to the collapse occurring slightly faster than in the fiducial M\_TD simulation. The increased amount of atomic hydrogen in the gas when starting from a 50:50 composition translates to increased production of molecules through hydrogenation in the ice, as evidenced by enhanced abundances of $\rm H_2CO$ and $\rm CH_3OH$ at large radii in simulations~M\_TD(50:50). However, the volume density in the outer regions is low, and hence the column density of $\rm CH_3OH$, for example, is higher in simulation~M\_TD(50:50) than in M\_TD by a factor of 1.8 only (for central $\rm H_2$ density of $10^5\,\rm cm^{-3}$). The relative ice abundances are also shifted when compared with the fiducial model (compare Figs.\,\ref{fig:abundances_iceratios}~and~\ref{fig:abundances_iceratios_halfAtomicH}). Using the 50:50 initial abundances leads to a better match of the CO ice column density with respect to observations and the $\rm CH_3OH/H_2O$ ratio also approaches the observed value in the very inner core, but the already overestimated production of species that form via barrierless hydrogenation ($\rm NH_3$, $\rm CH_4$) is made slightly worse by the increased availability of atomic hydrogen. We note that in any case the comparison to observations is not reliable because we are not attempting to model the exact physical conditions probed by the observations. Nevertheless, the results of this additional simulation reinforce the important role that initial conditions play in interstellar chemistry. 

\end{document}